\documentclass[pre,aps,showpacs,nofootinbib]{revtex4}
\usepackage{graphicx}
\usepackage{natbib}
\usepackage{bm}
\usepackage{amssymb}
\usepackage{amsmath}
\usepackage{euscript}
\usepackage{esint}
\usepackage{prelim2e}
\usepackage{slashed}

\begin{document}

\title{Infrared nullification of the effective electromagnetic field at finite temperature}

\author{Kirill A. Kazakov}
\author{Vladimir V. Nikitin}
\affiliation{%
Department of Theoretical Physics, Physics Faculty, Moscow State
University, 119899, Moscow, Russian Federation}

\begin{abstract}
The problem of infrared divergence of the effective electromagnetic field at finite temperature ($T$) is revisited. A model of single spatially localized electron interacting with thermal photons is considered in the limit $T\to 0$ using two different regularization schemes. The first is based on the shift $i0\to i\varepsilon$ of the electron propagator pole in the complex energy plane, and is used to explicitly calculate the effective field in the one-loop approximation. We show that the matrix-valued imaginary part of the electron self-energy can be consistently related to the pole shift, and that the presence of the heat bath leads to appearance of an effective $\varepsilon \sim T,$ thus providing a natural infrared regulator of the theory. We find that the one-loop effective Coulomb field calculated using this $\varepsilon$ vanishes. The other scheme combines an infrared momentum cutoff with smearing of the $\delta$-functions in the interaction vertices. We prove that this regularization admits factorization of the infrared contributions in multi-loop diagrams, and sum the corresponding infinite series. The effective electromagnetic field is found to vanish in this case too. An essentially perturbative nature of this result is emphasized and discussed in connection with  the long-range expansion of the effective field.
\end{abstract}
 \maketitle

\section{Introduction}

Interpretation of the effective (mean) field calculations is one of the most subtle questions in quantum field theory. It is raised every time one attempts to go beyond the realm of the scattering matrix, and is of vital importance in investigating such issues as the spontaneous symmetry breaking in the Electroweak theory, the quark-gluon plasma effects in quantum Chromodynamics, particle creation and inflation in quantum cosmology {\it etc.} In essential, this question consists of the following three apparently unrelated parts: 1) The gauge-dependence problem, {\it i.e.,} dependence of the effective field on the choice of the auxiliary conditions used to fix gauge invariance, 2) Measurability of the fields, {\it i.e.,} consistency of the formal field-theoretic predictions with the principal limitations of quantum measurements, and 3) The presence of infrared divergences.

The first two of these particular problems are formidable in Yang-Mills theories, which is why the question of interpretation of the effective field is still far from its resolution there. In quantum electrodynamics, on the contrary, they are relatively simple and amenable to general analysis. In fact, the proof of gauge independence of the effective quantities such as Heisenberg-Euler Lagrangian, the effective (mean) field, {\it etc.} is not difficult and has been given in quite general form \cite{equiv}. Furthermore, measurability of the electromagnetic field was proved in the fundamental works by Bohr and Rosenfeld \cite{bohr1,bohr2}. In particular, it was argued in \cite{bohr2} that the measuring apparatus can in principle be designed so as to take into account radiative corrections despite the presence of ultraviolet divergences. This consideration was subsequently generalized and carried over to the gravitational interaction in \cite{dewitt}.

What remains unsolved, however, is the problem of infrared divergences of the fields produced by elementary particles. Namely, it is known very well that the vertex formfactors of massive field quanta, used in constructing effective fields, are infrared-divergent. In this respect, electromagnetic interaction has many similarities with gravity, in particular, regarding resolution of the infrared catastrophe. Unfortunately, the corresponding proof \cite{bloch,leenaum} of cancelation of the infrared divergences in the S-matrix does not apply to the effective fields. In fact, the very statement of the problem in the two cases is quite different. The Bloch-Nordsieck theorem states that infrared divergences in the radiative corrections to the scattering cross sections exactly cancel those due to the emission of real soft photons. At the same time, the latter do not appear in the effective field formalism at all. An opinion can be found in the literature (see, {\it e.g.,} \cite{donoghue}) that this problem can be overcome by using the effective potential defined via the S-matrix according to \cite{toyoda,nambu}. It is known, however, that such definition is highly ambiguous \cite{kikugawa}: There are infinitely many potentials which lead to the same S-matrix, and not all of this arbitrariness can be traced to the gauge freedom in the theory. As a result, even to achieve agreement with the classical theory requires introduction of certain prescriptions at each order of the perturbation expansion. Apart from this difficulty, such a definition would not help investigating the issues mentioned at the beginning.

By this reason, the effective field formalism can be applied, strictly speaking, only to the fields produced by classical sources. If, for instance, the mass of a charged particle is sufficiently large, radiative corrections to its interaction with the electromagnetic field can be neglected, thus putting the question about their divergence aside. In fact, assumptions of this kind underlie the classic calculation \cite{serber} and its generalizations \cite{bialb}. However, this limitation looks rather unsatisfactory from the theoretical point of view, especially in the light of the positively resolved issue 2) above.

This problem is sharpened at finite temperatures, because the heat-bath effects introduce new low-energy singularities to the photon propagator. These additional singularities are generally believed to worsen infrared properties of the Feynman integrals. Previous investigations of the infrared problem have been aimed mainly at generalizing the Bloch-Nordsieck and Kinoshita-Lee-Nauenberg theorems to nonzero temperatures \cite{tryon,donoghue1985,manjavidze,altherr,weldon,indumathi,muller}. It should be mentioned in this connection that extending the S-matrix formalism to finite temperatures is itself a nontrivial task. Some general results in this direction were obtained in \cite{manjavidze}, where an S-matrix connecting states with different temperature was constructed. An independent definition of the finite-temperature S-matrix is given in \cite{weldon} where infrared finiteness of the S-matrix is proved in the eikonal approximation. These results were subsequently improved in \cite{indumathi}. Though in a different way, cancelation of the lowest-order infrared divergences at finite temperature was earlier demonstrated in \cite{tryon}. However, it was argued later that the infrared divergences treated either way still persist at arbitrary temperatures \cite{muller}. It is important to stress that all these developments proceed within the conventional approach based on summation of the scattering cross sections over suitable classes of processes. As was already mentioned, their results do not apply to the effective fields.

There is an argument often used in the literature, according to which infrared divergences in the formfactors can be discarded within the long-range expansion of the electromagnetic or gravitational fields. Indeed, it is known that the infrared-divergent contributions considered as functions of the momentum transfer, $\bm{p}$ (which is the variable Fourier-conjugated to the distance from the field source), are analytic at $\bm{p} = 0.$ Though infinite, such contributions  consequently give rise to terms proportional to the Dirac delta-function or its derivatives in the coordinate space. The divergent part of the effective field thus turns out to be confined to the region of source localization. Yet, this argument cannot be recognized convincing. The point is that such terms appear at every order of the perturbation theory, and each order adds new derivatives acting on the delta-function. Therefore, the question as to locality of the divergent contribution can be settled only after summing the infinite series of differential operators, because the sum of local operators may well turn out to be a nonlocal operator.

Turning back to the infrared problem in the general case, we would like to stress that inclusion of the finite temperature effects into consideration is a question of principle: no matter how small temperature might be, it is never zero exactly, and the question as to whether $T\approx 0$ can be replaced by $T=0$ can be answered only by investigating the general nonzero-temperature case. In other words, continuity of this sort, being a necessary physical requirement for the very possibility to neglect the heat-bath effects, is to be proved rather than postulated. It may well turn out that in view of the aforementioned principal inequality $T\ne 0,$ the notion of the vacuum effective field is to be defined as the limit of its finite-temperature counterpart for $T\to 0.$ The purpose of the present paper is to examine the infrared problem from this point of view. In this respect, our investigation differs from the standard approach based on the high-temperature expansion, which is commonly used in actual calculations (consideration of the low-temperature case can be found, {\it e.g.,} in \cite{donoghue1985,muller}). We restrict ourselves to the electromagnetic interaction because its diagrammatics is much simpler than gravitational, and admits in certain cases partial summation of the perturbation series, which cannot be performed in the presence of self-interaction of the gauge field. We introduce a simplified model for a single spatially-localized charge interacting with the heat-bath photons at equilibrium, which neglects the influence of the photon interactions on their statistical distribution. This simplification is in fact a valid approximation in investigating the infrared problem, because it is the low-energy asymptotic of the photon distribution that is only important in this case, while the radiative corrections to this distribution vanish at zero momentum. This model requires some modification of the diagrammatic rules, which is described in Sec.~\ref{prelim}.
To examine the infrared divergences of the effective electromagnetic field, we use two different regularization schemes. The first, called $\varepsilon$-regularization for brevity, consists in the usual finite shift $i0 \to i\varepsilon$ of the poles of the electron propagator in the complex energy plane. This regularization is quite natural in view of the fact that the temperature effects lead to appearance of an imaginary part in the electron self-energy, thus providing an effective infrared regulator which replaces $\varepsilon$ by a quantity $\sim T.$ This scheme is also very convenient in actual calculations as it reveals some important features hidden in other regularizations, such as the cancelation of counterterm contributions to the effective field, considered in Sec.~\ref{oneloop}. However, it does not admit factorization of many-loop infrared contributions, which hinders partial summation of the perturbation series. The other scheme, called $\lambda$-regularization in what follows, is based on the introduction of an infrared cutoff, and is described in Sec.~\ref{lambdareg}. This scheme does admit summation of the infrared contributions similar to that employed in the standard treatment of the infrared catastrophe. This summation is performed in Sec.~\ref{factorization}. The general conclusion drawn in Sec.~\ref{conclude} is that both regularizations suggest vanishing of the effective electromagnetic field calculated using the perturbation theory, thus revealing an essentially non-perturbative nature of the infrared problem. The paper has an Appendix where some technical aspects of our work are discussed on the basis of the Ward identities.

\section{Preliminaries}\label{prelim}

Consider the electromagnetic field produced by an electron of mass $m,$ which is on average at rest and interacts with the virtual as well as real photons in equilibrium at finite temperature\footnote{We use relativistic units $\hbar = c = 1.$ Also, the Minkowski metric is $\eta_{\mu\nu} = {\rm diag}\{+1,-1,-1,-1\}.$} $T \ll m.$ Neglecting the effects related to the quantum spreading of the electron wavefunction, the field is purely electric, so that it is sufficient to find the effective scalar potential
\begin{eqnarray}\label{eff}&&
A_0^{\rm eff} (x) = \EuScript{N}^{-1}{\rm Tr} \left( A_0 (x)e^{- \beta H_\phi } \varrho\right)\,, \quad \EuScript{N} = {\rm Tr} \left( e^{- \beta H_\phi } \varrho\right)\,, \quad \beta = 1/T\,.
\end{eqnarray}
\noindent Here $ A_0 (x)$ is the Heisenberg picture operator of the scalar potential, $H_\phi$ the Hamiltonian of free photons, $\varrho$ the electron density matrix, and the trace is over all photon states as well as the single electron states. The matrix $\varrho$ is assumed to describe a non-thermalized state, {\it viz.}, a state in which the electron is spatially localized. Having chosen $H_\phi$ to be the free photon Hamiltonian we thereby omit corrections to the photon distribution due to their interactions. It is readily seen that within the one-loop approximation, these interactions do not change the effective field indeed. In fact, in investigating the infrared divergence problem at higher orders they can be neglected as well. This is because in this case, it is the low-energy asymptotic of the photon distribution that only matters, while the radiative corrections to the photon Green functions vanish at zero momentum to all orders. Also, the electron Hamiltonian might have been added to the exponent in Eq.~(\ref{eff}). However, since there is only one nonrelativistic electron, and $T\ll m,$ its contribution can be easily shown to factorize and cancel the corresponding contribution to the normalization factor $\EuScript{N},$ leaving the value of the effective field unchanged.

\begin{figure}[h]
\begin{center}
     \includegraphics[width=9cm]{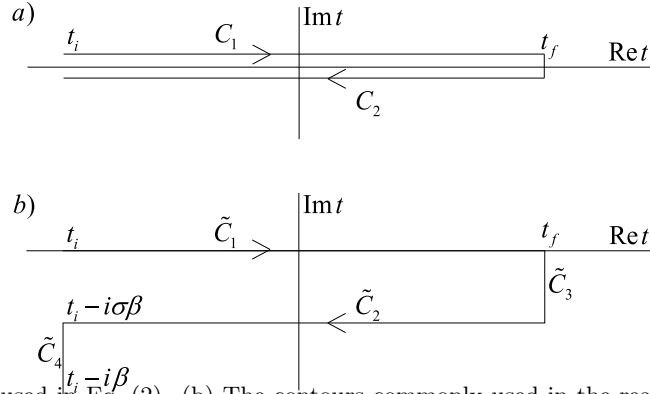}
\caption{(a) Integration contour used in Eq.~(\ref{effInt}). (b) The contours commonly used in the real-time techniques, parameterized by $\sigma \in (0,1).$ The limit $t_i\to-\infty, t_f\to\infty$ is understood.}
\label{contour}
\end{center}
\end{figure}

To evaluate the right hand side of Eq.~(\ref{eff}), we shall use the general framework of the real time approach \cite{landsman}. Since there is only one charged particle present which is not in thermal equilibrium with the photon bath, the standard finite-temperature-field-theory techniques are not directly applicable to the calculation of $A_0^{\rm eff}.$ The relevant extension of the real time formulation can be obtained as follows. According to the Schwinger-Keldysh method \cite{Keldysh,Schwinger}, the expectation value (\ref{eff}) can be written as
\begin{eqnarray}\label{effInt}&&
A_0^{\rm eff} (x) = \EuScript{N}^{-1}{\rm Tr} \left( T_c \left[ {\exp i\left( {\int\limits_C {d^4 x} L_I (x)} \right) A_0 (x)} \right]_{in} e^{ - \beta  H_\phi  } \varrho\right)\,,
\end{eqnarray}
where $L_I$ is the interaction Lagrangian, and the $x^0$-integration is along the contour $C=C_1\cup C_2$ in the complex time plane, shown in Fig.~\ref{contour}(a). The subscript $in$ indicates that the operators are taken in the interaction picture; $T_c$ is the usual ordering along the contour $C.$ Concerning the choice of the time contour, the following circumstance must be mentioned from the outset. It is a common procedure of the real time techniques that the factor $e^{- \beta H},$ with $H$ the full Hamiltonian, is promoted into an integral along the interval $[t_i,t_i-i\beta]$ in the complex time plane, followed by a contour deformation shown in Fig.~\ref{contour}(b). Taking the limit $t_i\to-\infty, t_f\to\infty$ then allows one to get rid of the contributions of the contour segments $\tilde{C}_3$ or $\tilde{C}_4.$ This procedure presupposes eventual factorization of such contributions, resulting from the vanishing of the two-point Green functions having one argument on $\tilde{C}_{3}$ or $\tilde{C}_4$ and the other on $\tilde{C}_1$ or $\tilde{C}_2.$ It is known, however, that in theories containing massless particles, this transformation is jeopardized by the infrared singularities \cite{landsman}. It turns out that the model we consider is the case where the trick with the contour deformation is not legitimate. Specifically, an explicit calculation shows that the one-loop diagrams involving the Green functions which connect $\tilde{C}_{1,2}$ with $\tilde{C}_3$ actually do not vanish. In particular, omitting them makes the effective electromagnetic field  complex-valued. On the other hand, the replacement of $H$ by the free photon Hamiltonian makes it unnecessary to follow the above-mentioned procedure. Thus, to avoid the complications associated with the contour deformations, we adhere to the original Schwinger-Keldysh formulation.

Perturbation expansion of the right hand side of Eq.~(\ref{effInt}) leads to the sum of terms of the form
\begin{eqnarray}\label{expand}
\int {d^4 y} \int {d^4 y'} ...{\rm Tr}\left(e^{ - \beta H_\phi  } T_c \left[A_0 (x) A_\mu  (y) A_\nu  (y')...\right]\right)  \nonumber \\
 \times {\rm Tr}\left(T_c [:\bar \psi (y)\gamma ^\mu  \psi (y)::\bar \psi (y')\gamma ^\nu  \psi (y'):...]\varrho\right).
\end{eqnarray}
Hereon all the field operators are taken in the interaction picture, and the subscript $in$ is omitted. As usual, the particle propagators acquire $2\times2$ matrix structure as the result of $T_c$-ordering. By virtue of the Wick theorem, the first factor in the above expression is represented as the sum of products of various finite-temperature photon two-point functions, while the second a similar sum constructed from the vacuum Green functions for the electron-positron field, and the single factor ${\rm Tr}\left(\bar \psi (y')\psi (y'')\varrho\right)$.
Each propagator bears indices $1$ or $2$ assigned to its ends, according to whether the end belongs to the contour branch $C_1$ or $C_2,$ and connects vertices of the corresponding type, with an extra minus sign for each 2-type vertex. The observation point $x$ is assigned index 1. In momentum space, the matrix photon propagator in the Feynman gauge reads $\mathfrak{D}_{\mu\nu}(k) = \eta_{\mu\nu}D^{(ij)}(k),$ where
\begin{eqnarray}\label{phiphot}
D^{(11)}(k) &=& - \tilde D^{(22)}(k) = \left[\frac{1}{k^2 + i0} - {2\pi i
n(\bm{k})\delta(k^2)}\,\right], \nonumber \\
D^{(21)}(k) &=& - 2\pi i \left[\theta(k_0)+n(\bm{k})\,\right]\delta(k^2), \nonumber \\
D^{(12)}(k) &=& - 2\pi i \left[\theta(-k_0)+n(\bm{k})\,\right]\delta(k^2)\,,
\quad n(\bm{k})=\frac{1}{e^{\beta|\bm{k}|} - 1},\nonumber
\end{eqnarray}
\noindent while the electron matrix propagator $\mathfrak{D}_{e}(k)$ has the elements
\begin{eqnarray}\label{electronpr1}
D_e^{(11)} (k) &=&  - \tilde D_e^{(22)} (k) = \frac{{\slashed{k} + m}}{{m^2  - k^2  - i0}}\,, \\
D_e^{(21)} (k) &=&  2\pi i\delta(k^2  - m^2 ) \theta (k_0 )(\slashed{k} + m)\,, \\
D_e^{(12)} (k) &=&  - 2\pi i\delta(k^2  - m^2 ) \theta (-k_0 )(\slashed{k} + m)\,,\label{electronpr3}
\end{eqnarray}\noindent where the tilde symbolizes the special operation of complex conjugation which treats the $\gamma$-matrices as real quantities: $\tilde{\gamma} = \gamma.$ It will be convenient to represent $\mathfrak{D}_{e}(k)$ in the form
\begin{eqnarray}\label{electronpr2}
\mathfrak{D}_{e}(k) &=&
M(k)\left(\begin{array}{cc}
D_F & 0\\
0 & -\tilde D_F
\end{array}\right)M(k), \quad D_F(k) = \frac{1}{m^2-k^2-i0}, \nonumber\\
M(k) &=& \left(\begin{array}{cc}
1 & -\theta(-k_0) \\
\theta(k_0) & 1 \label{M}
\end{array}\right)\,,
\end{eqnarray}\noindent which is established directly using Eqs.~(\ref{electronpr1})--(\ref{electronpr3}).

\section{One-loop contribution to the effective field}\label{oneloop}

\subsection{$\varepsilon$-regularization}\label{deltareg}

Infrared singularities in the electromagnetic formfactors and self-energy contributions to the effective field, as well as ultraviolet divergences require introduction of intermediate regularization. The simplest and very convenient choice we make in order to investigate the effective field in the one-loop approximation is
the usual finite shift of the electron propagator poles in the complex $k^0$-plane. Namely, we replace the function $D_F(k)$ in Eq.~(\ref{electronpr2}) by
$$D_F(k) = \frac{1}{m^2-k^2-i\varepsilon}, \quad \varepsilon>0$$
(similar regularization of the photon propagator is unnecessary). Accordingly, the $\delta$-function appearing in the electron propagator is regularized as
$$2\pi i\delta(k^2  - m^2) \to \frac{1}{m^2 - k^2 - i\varepsilon} - \frac{1}{m^2 - k^2 + i\varepsilon}\,.$$ Obviously, $\varepsilon$ plays the role of infrared regulator, and for brevity, we call this scheme $\varepsilon$-regularization in what follows. In fact, the introduction of $\varepsilon$-regulator is necessary in the real-time formulation of thermal field theory in order to correctly treat the products of singular functions appearing in the Feynman diagrams. As we shall see, it helps revealing some nontrivial features of the one-loop contribution to the effective field, which are hidden in other regularizations. To make the $\varepsilon$-regularization physically meaningful, one should impose some restriction on the value of $\varepsilon$ with respect to the physical parameters of the theory. Since $\varepsilon$ has dimension $[mass]^2,$ such a restriction should be smallness of $\varepsilon$ in comparison with some quadratic combination of $T$ and $m.$ We assume throughout that $\varepsilon$ satisfies
\begin{eqnarray}\label{deltacond}
\varepsilon \ll m T.
\end{eqnarray}
\noindent
It will be seen below that this condition appears naturally in the theory. In addition to that, we introduce a momentum threshold, $\Lambda,$ separating the infrared effects. It satisfies $T\ll \Lambda\ll m,$ and identifies the photons with $k < \Lambda$ as ``soft.'' As to the usual ultraviolet divergences, they are supposed to be regularized using some conventional means, say, the dimensional technique.

\begin{figure}[h]
\begin{center}
     \includegraphics{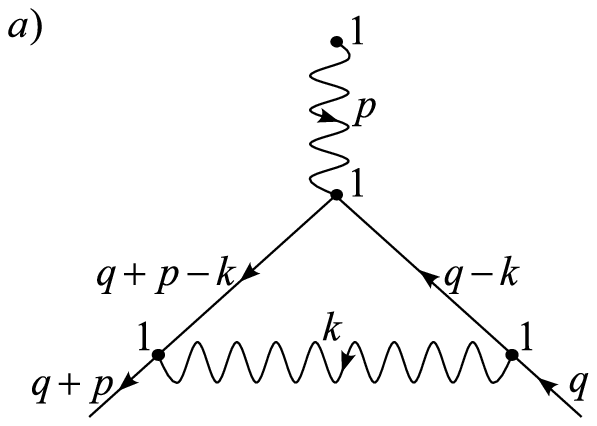}
     \includegraphics{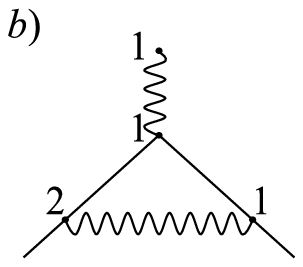}
     \includegraphics{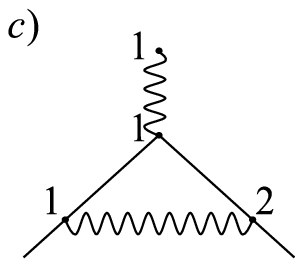}
     \includegraphics{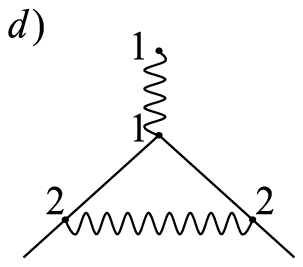}
\caption{Vertex contribution to the effective field.}
\label{Diag}
\end{center}
\end{figure}

\begin{figure}[h]
\begin{center}
     \includegraphics{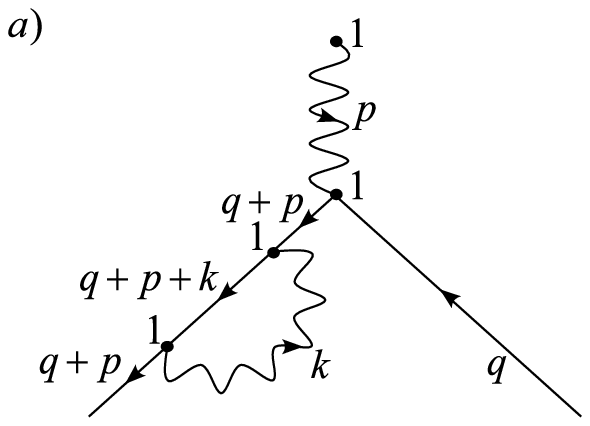}
     \includegraphics{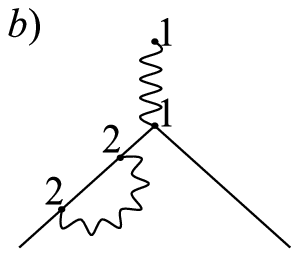}
     \includegraphics{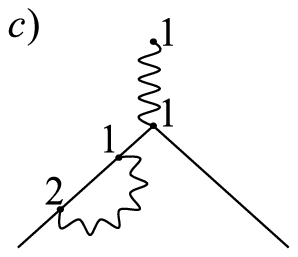}
     \includegraphics{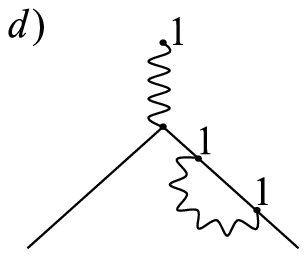}
\caption{Electron self-energy contribution to the effective field.}
\label{Self}
\end{center}
\end{figure}

\begin{figure}[h]
\begin{center}
     \includegraphics{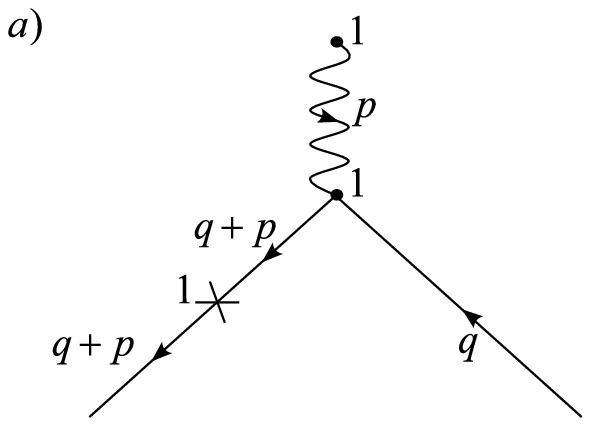}
     \includegraphics{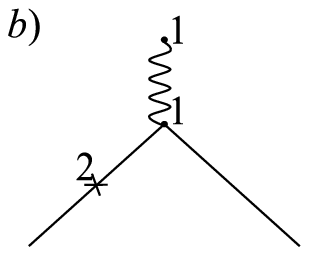}
     \includegraphics{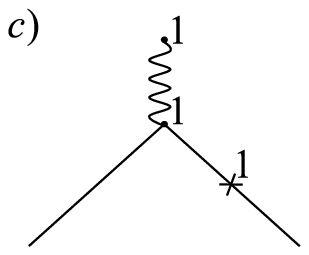}
\caption{The self-energy counterterm diagrams.}
\label{Kontr}
\end{center}
\end{figure}

Upon transition to the momentum space, $A^{\rm eff}_0(x)$ takes the form
\begin{eqnarray}\label{Fur_e}
A^{\rm eff}_{0}(x) = \sum\limits_{\sigma ,\sigma '}\iint
\frac{d^3 \bm{q}}{(2\pi)^3} \frac{d^3\bm{p}}{(2\pi)^3}e^{i px}D^{(11)}(p) \varrho_{\sigma \sigma'}(\bm{q},\bm{q} + \bm{p})\bar{u}_{\sigma'}(\bm{q}+\bm{p})J(p,q)u_{\sigma}(\bm{q})\,,
\end{eqnarray}
\noindent where the matrix-valued function $J(p,q)$ is essentially a sum of products of several momentum-space propagators. In this formula, the values of $q^0$ and $p^0$ are fixed by the mass-shell condition ${q^0 = \sqrt{\bm{q}^2 + m^2}>0}$ and the energy-momentum conservation. Since the field-producing charge is non-relativistic, ${p^0 = (\bm{q}+\bm{p})^2/2m - \bm{q}^2/2m\,.}$ The bispinor amplitudes and the momentum-space density matrix are normalized on unity: $$\bar{u}_{\sigma}u_{\sigma}=1, \quad \sum\limits_{\sigma}\int
\frac{d^3 \bm{q}}{(2\pi)^3}\varrho_{\sigma \sigma}(\bm{q},\bm{q}) = 1\,.$$
It is easily verified that in the tree approximation,
\begin{eqnarray}\label{tree}
J(p,q) = - e\gamma^0.
\end{eqnarray}
\noindent
The one-loop diagrams representing $A^{\rm eff}_0$ are shown in Figs.~\ref{Diag}--\ref{Kontr}.

\begin{figure}[h]
\begin{center}
     \includegraphics{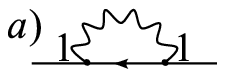}
     \includegraphics{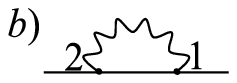}
     \includegraphics{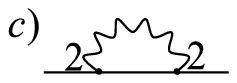}
\caption{One-loop contribution to the normalization factor $\EuScript{N}.$}
\label{nfactor}
\end{center}
\end{figure}

It is not difficult to see that the diagrams (c) and (d) in Fig.~\ref{Diag} are irrelevant. This is because they involve the function $D_e^{(12)}(q-k)\sim\theta (k_0-q_0),$ and therefore do not contribute at small loop momenta. By this reason, diagrams involving this function are omitted in Figs.~\ref{Self}, \ref{Kontr}.
Less trivial is the fact that the counterterm diagrams do not contribute either (the procedure of introducing temperature-dependent counterterms is discussed in detail {\it e.g.}, in \cite{semenoff1985,keil1989}). For the proof, we note that the vertices in Fig.~\ref{Kontr}, generated by the counterterm  Lagrangian, are contracted with the bispinor amplitudes on the mass shell. Therefore, the corresponding vertex factor is proportional to the unit matrix. Denoting $l_c$ the proportionality coefficient, we find for the sum of diagrams in Fig.~\ref{Kontr}
\begin{eqnarray}\label{ctcancel}
&&J_{(4)}(p,q) = ie l_c
\left[ {\frac{{\slashed{q} + \slashed{p} + m}}{{ - i\varepsilon }}\gamma^0
- (\slashed{q} + \slashed{p} + m)\left(\frac{1}{{ - i\varepsilon }} - \frac{1}{{i\varepsilon }}\right)\gamma^0
+ \gamma^0  \frac{{\slashed{q} + m}}{{ - i\varepsilon }}  } \right] = 0,
\end{eqnarray}\noindent since $(\slashed{q} + \slashed{p})\gamma^0 \to m\gamma^0,$ $\gamma^0\slashed{q}\to m\gamma^0$ upon contraction with $\bar{u},u$ on the mass shell. We can interpret this fact by saying that when calculating the effective field, the usual renormalization prescriptions for the electron propagator can be assumed fulfilled automatically. Thus, only diagrams in Fig.~\ref{Diag}(a),\ref{Diag}(b) and Fig.~\ref{Self}(a)--\ref{Self}(d) remain to be considered.

Our primary aim below is to extract the small loop-momentum parts of the above diagrams, and to show that the one-loop contribution to the effective field is infrared-finite. In doing this, we consider arbitrarily small but finite temperatures, and calculate asymptotics of the diagrams as $T \to 0.$
The calculation reveals two types of infrared-divergent terms in these asymptotics. Their leading at $T\to 0$ terms are proportional to $\ln(\beta\varepsilon),$ while the subleading ones are linear in temperature. It turns out that the singular part $\ln\varepsilon$ of the former exactly cancels the corresponding vacuum contribution, so that the finite remainder proportional to $\ln\beta$ determines the low-temperature behavior of the vertex and the electron self-energy contributions to the effective field. The role of the other terms is different. The point is that the linear in $T$ contribution to the electron self-energy contains an imaginary part, and acts as a natural infrared regulator of the theory. In other words, taking into account the subleading terms is equivalent to replacing $\varepsilon \to {\rm const}\cdot T.$  It turns out that this gives rise to some finite temperature-independent terms in the vertex formfactor which exactly cancel the tree contribution.

\subsection{Electron self-energy contribution}

To begin with, let us show that the one-loop contribution to the normalization factor $\EuScript{N}$ vanishes. This is conveniently done using the following representation for the matrix electron self-energy \cite{niemi}
\begin{eqnarray}\label{Sform}
\Sigma(q) =
M^{-1}(q)\left(\begin{array}{cc}
\Sigma_F & 0\\
0 & -\tilde \Sigma_F
\end{array}\right)\,M^{-1}(q)
=\left(\begin{array}{cc}
\Sigma_F & \theta(-q_0)(\Sigma_F - \tilde\Sigma_F) \\
-\theta(q_0)(\Sigma_F - \tilde\Sigma_F) & -\tilde \Sigma_F
\end{array}\right)\,,
\end{eqnarray}
where $M$ is defined in Eq.~(\ref{M}), and $\Sigma_F \equiv \Sigma ^{(11)}(q).$ It must be mentioned that this representation is generally valid only in the limit $\varepsilon\to 0.$ This is because its derivation is based on the identity $\EuScript{D}_e^{(11)} + \EuScript{D}_e^{(22)} = \EuScript{D}_e^{(12)} + \EuScript{D}_e^{(21)}$ for the exact electron propagator, which is a direct consequence of the time-ordering prescription, and hence, is affected by the $\varepsilon$-regularization. However, explicit calculations show that it holds true also for the leading terms we are interested in, and so we use Eq.~(\ref{Sform}) below to render the formulas more compact. The one-loop contribution to $\EuScript{N},$ shown in Fig.~\ref{nfactor}, reads
\begin{eqnarray}
\EuScript{N}^{\rm loop} = \sum\limits_{\sigma ,\sigma '}\int
\frac{d^3 \bm{q}}{(2\pi)^3} \varrho_{\sigma \sigma'}(\bm{q},\bm{q})\bar{u}_{\sigma'}(\bm{q})\left(\Sigma^{(11)}(q) + \Sigma^{(21)}(q) + \Sigma^{(22)}(q)\right)u_{\sigma}(\bm{q})\,.\nonumber
\end{eqnarray}
Using Eq.~(\ref{Sform}) shows immediately that the integrand is zero identically (in this case, the $\varepsilon$-regularization is unnecessary).

Moreover, it turns out that the real Lorentz-invariant parts of the electron self-energy (reality being understood in the sense of the operation $\widetilde{\ \ }$\ ) also cancel upon substitution to the effective field. To see this, let us write down the contributions of diagrams in Fig.~\ref{Self} to the function $J(p,q)$
\begin{eqnarray}\label{j3a}
J_{(3a)}&=&
ie\Sigma _F (q+p)\frac{1}{\varepsilon}(\slashed{q} + \slashed{p} + m)\gamma ^0 \label{FormTo1}\\
J_{(3b)}&=&
-ie\tilde\Sigma _F(q+p) \frac{2}{\varepsilon }(\slashed{q} + \slashed{p} + m)\gamma^0  \label{FormTo2}\\
J_{(3c)}&=&
ie\left\{\tilde\Sigma _F(q+p) - \Sigma _F(q+p)\right\} \frac{1}{\varepsilon }(\slashed{q} + \slashed{p} + m)\gamma^0 \label{FormTo3}\\
J_{(3d)}&=&
ie \gamma^0(\slashed{q} + m) \frac{1}{\varepsilon } \Sigma _F (q)\label{FormTo4}
\end{eqnarray}\noindent Recalling that these expressions are sandwiched between bispinor amplitudes, we may replace $\slashed{q} + \slashed{p}\to m$ in the first three of these expressions, and $\slashed{q}\to m$ in the last one. Taking into account also the mass-shell conditions $q^2 = (q+p)^2 = m^2,$ we conclude that the Lorentz-invariant part of their sum involves only the difference $\tilde\Sigma _F - \Sigma _F$. It follows that the vacuum contribution is $O(\varepsilon),$ since it is Lorentz-invariant and real for $\varepsilon=0.$

The explicit expression for $\Sigma_F$ at the one-loop level reads
\begin{eqnarray}
\Sigma _F (q) = ie^2 \int {\frac{{d^4 k}}{{(2\pi )^4 }}\frac{{\gamma ^\mu  (\slashed{q} + \slashed{k} + m)\gamma_\mu  }}{{m^2  - (q + k)^2  - i\varepsilon }}\left[ {\frac{1}{{k^2  + i0}} - 2\pi in(\bm{k})\delta(k^2 )} \right]}\,.
\end{eqnarray}\noindent In view of what was just said, the ultraviolet divergence of $\Sigma_F$ is of no importance for the present computation. Furthermore, since its imaginary part comes from integration over small $k,$ we may cut the integral at $k = \Lambda \gg T.$ As $T$ is arbitrarily small, $\Lambda$ can be taken small enough to neglect $\slashed{k}$ in the numerator as well as $k^2$ in the denominator. Then the vacuum contribution (corresponding to the first term in the square brackets) takes the form
[we have to retain the $O(\varepsilon)$-terms because they are combined with the mass-shell electron propagators in Fig.~\ref{Self}, which are $O(1/\varepsilon)$]
\begin{eqnarray}
\Sigma _F ^{\rm vac}(q) &=& - ie^2 (2m - \slashed{q})\int\limits^{\Lambda} \frac{{d^4 k}}{{(2\pi )^4 }}\frac{1}{{2kq + i\varepsilon }}\frac{1}{{k^2  + i0}} + \cdots \nonumber\\ &=& \frac{e^2}{2 \pi^2} (2m - \slashed{q})\int \frac{d\Omega}{4\pi}
\int\limits_0^\Lambda \frac{\kappa d\kappa}{2\kappa a - i\varepsilon} + \cdots
= - i\frac{e^2}{8\pi^2}(2m - \slashed{q})\frac{\varepsilon}{m^2} \ln\frac{\varepsilon}{m^2} + \cdots, \nonumber
\end{eqnarray}
where $a = \hat{k}q,$ $\hat{k}_0=|\bm{\hat{k}}|=1,$ $\int d\Omega$ is the integration over directions of $\bm{k},$ and ``$\cdots$'' stands for real terms, or terms of higher order in $T,\varepsilon.$

To find the heat-bath contribution, let us use the condition (\ref{deltacond}) to introduce an auxiliary intermediate scale $\lambda$ such that $\varepsilon/m \ll \lambda \ll T.$ Then the loop integral can be evaluated as
\begin{eqnarray}\label{SigHeat}&&
\hspace{-0,8cm}\Sigma _F^{\rm heat}(q) = \frac{e^2}{2 \pi^2} (2m - \slashed{q})
\int \frac{d\Omega}{4\pi}
\int\limits_0^\infty n(\kappa) \kappa d\kappa \left(\frac{1}{2\kappa a-i\varepsilon}-\frac{1}{2\kappa a+i\varepsilon}\right)  \nonumber \\
&=& \frac{e^2}{2 \pi^2} (2m - \slashed{q})
\int \frac{d\Omega}{4\pi} \left[
\int\limits_0^\lambda d\kappa\left(\frac{1}{\beta}-\frac{\kappa}{2} \right)  \left(\frac{1}{2\kappa a-i\varepsilon}-\frac{1}{2\kappa a+i\varepsilon}\right)
+ \frac{i\varepsilon}{2 a^2}\int\limits_\lambda^\Lambda \frac{n(\kappa)}{\kappa}d\kappa
\right ] + \cdots  \nonumber \\
&=& i\frac{e^2}{8 \pi^2} (2m - \slashed{q})
\left[
\frac{\varepsilon}{m^2}\ln\frac{\beta\varepsilon}{m} + \frac{2\pi}{\beta}\int \frac{d\Omega}{4\pi}\frac{1}{a}
\right ] + \cdots, \nonumber
\end{eqnarray}\noindent Adding the two contributions yields
$$\Sigma _F (q) = i\frac{{e^2 }}{8\pi ^2}(2m - \slashed{q})\left[\frac{\varepsilon}{m^2}\ln(\beta m)  + \frac{2\pi}{\beta}\langle a^{-1}\rangle\right] + \cdots,$$
where the angular brackets denote averaging over directions of the 3-momentum, $$\langle f\rangle = \int \frac{d\Omega}{4\pi}f\,.$$ Using the condition $|\bm{q}|\ll m,$ this expression can be further simplified to
\begin{eqnarray}\label{sigma}
\Sigma _F (q) &=& i\frac{{e^2 }}{8\pi ^2}(2m - \slashed{q})\left[\frac{\varepsilon}{m^2}\ln(\beta m)  + \frac{2\pi}{\beta m}\right] + \cdots,
\end{eqnarray}\noindent which will be sufficient for our purposes below. We note, first of all, that the heat-bath contribution proportional to $\ln\varepsilon$ cancels that of the vacuum part. This fact allows us to use $\Sigma_F$ to construct a function of $\varepsilon,$ regular in a vicinity of $\varepsilon =0.$  Indeed, $\Sigma_F$ was defined initially for $\varepsilon>0,$ and the found cancelation of logarithmic singularities means that it can be continued analytically to the region $\varepsilon\leqslant 0,$ and that this continuation is unique. Let us denote it $\Sigma^{(+)}_F = \Sigma^{(+)}_F(q,\varepsilon).$ We need this continuation because  Eq.~(\ref{sigma}) shows that at finite temperature, the electron mass acquires an imaginary part $\sim \alpha T$ ($\alpha = e^2/4\pi$ is the fine structure constant), and this part is positive.\footnote{The fact that the mass eigenvalues in thermal field theory can have positive as well as negative imaginary part was emphasized in \cite{semenoff1985,landsman1991}.} This means in turn that its mass squared becomes $m^2 + i E,$ where $m$ is the (renormalized) vacuum mass, and $ E \sim \alpha mT.$
In other words, the finite-temperature effects introduce natural infrared regulator $ E,$ replacing in effect $\varepsilon\to \varepsilon -  E,$ so that removal of the infrared regularization is now to be understood as $\varepsilon\to -  E$ in the analytically continued functions. Specifically, the electron self-energy becomes in this limit $\Sigma^{(+)}_F(q,- E),$ while the mass-shell electron propagator, $(\slashed{q}+m)/(i E).$

In order to determine the precise value of $E,$ we have to relate the matrix-valued imaginary part of the electron self-energy to the scalar shift of the propagator pole. Since a change of $E$ changes the electron self-energy, which in turn affects $E$ itself, one has to use Eq.~(\ref{sigma}) to find the fixed point for $E,$ {\it i.e.,} the value which upon replacing $\varepsilon\to - E$ in the electron propagator would give the same $E$ as read off from the electron self-energy calculated using this propagator. Let us write, for brevity, ${\rm Im}\,\Sigma_F(q) = i\sigma_1 \slashed{q} + i\sigma_2 m,$ where $\sigma_{1,2} = O(\alpha)$ are real quantities whose explicit expressions can be found by replacing $\varepsilon \to - E$ in Eq.~(\ref{sigma}). Using the fact that ${\rm Re}\,\Sigma_F(q)$ vanishes on the mass shell (this condition is satisfied automatically, Cf. Sec.~\ref{deltareg}), we have for the electron propagator near its pole
$$\frac{1}{m-\slashed{q}+\Sigma_F(q)}
=\frac{\slashed{q} + m + i m\sigma_2 - i\slashed{q}\sigma_1}{m^2 - q^2 + 2i(q^2 \sigma_1 + m^2\sigma_2) + q^2\sigma^2_1 - m^2 \sigma^2_2}\,.$$
Since the leading terms in $\sigma_{1,2}$ are $O(\alpha),$ and do not contain $\ln(\beta m),$ within the accuracy of our calculation this expression reduces on the mass shell ($q^2 = m^2$) to $$\frac{\slashed{q} + m}{2im^2(\sigma_1 + \sigma_2)}\,.$$ On the other hand, according to the definition of $ E,$ the mass-shell electron propagator is
$$\frac{\slashed{q} + m}{i E}.$$ Comparison of the two formulas yields
$ E = 2m^2(\sigma_1 + \sigma_2),$ or, using Eq.~(\ref{sigma}),
$$ E = \frac{e^2 m^2}{4\pi^2}\left[\frac{- E}{m^2}\ln(\beta m)  + \frac{2\pi}{\beta m}\right].$$
Thus, we find
\begin{equation}\label{ZnachDelta}
 E=\frac{e^2 m}{2\pi\beta}\left(1+\frac{e^2}{4\pi^2}\ln(\beta m)\right)^{-1}\,.
\end{equation}\noindent It should be emphasized that the temperature in this calculation does not have to satisfy $e^2 \ln(\beta m) \ll 1,$ so that Eq.~(\ref{ZnachDelta}) is valid for all $T$ down to zero. Finally, using this result in the sum of the four contributions (\ref{j3a}) -- (\ref{FormTo4}), we arrive at the following expression for the electron self-energy contribution to the function $J(p,q)$
$$J_{(3)}(p,q) = e\gamma^0\frac{4m}{ E}{\rm Im}\,\Sigma^{(+)}_F = 2e\gamma^0.$$
Because of the mixing of orders caused by $ E = O(e^2)$ in the denominator, this contribution turns out to be a first-order quantity.

\subsection{Vertex contribution and nullification of the one-loop Coulomb potential}\label{vertex}

At last, let us consider diagrams in Fig.~\ref{Diag}. To extract the leading term as $T\to 0,$ we again introduce the cutoff $\Lambda,$ and replace $m^2  - (q - k)^2 \to 2kq$ in the mass-shell electron propagators. Performing the standard $\gamma$-matrix algebra with the help of the formula $(\slashed{q} + m)\gamma ^\mu  u_\sigma(\bm{q})= 2q^\mu u_\sigma(\bm{q}),$ one then finds that the infrared contribution of diagrams (a), (b) in Fig.~\ref{Diag} is contained in the expression
\begin{eqnarray}
J_{(2)} &=&
4ie^3\gamma^0 (m^2-p^2/2)\int\limits^{\Lambda} {\frac{{d^4 k}}{{(2\pi )^4 }}} \left[ \frac{1}{k^2  + i0}\frac{1}{2kq + i \varepsilon }\frac{1}{2k(q + p) + i \varepsilon } \right. \nonumber\\&& - \frac{1}{2kq - i \varepsilon }\frac{2\pi in(\bm{k})\delta(k^2 )}{2k(q + p) + i\varepsilon }
\left. { + \frac{2\pi i\delta(k^2 )\theta (k_0 )}{{2kq - i\varepsilon }}\left( {\frac{1}{{2k(q + p) - i \varepsilon }} - \frac{1}{{2k(q + p) + i\varepsilon }}} \right)} \right].\nonumber
\end{eqnarray}
Integration over $k_0$ yields
\begin{eqnarray}
J_{(2)} =
 \frac{e^3\gamma^0 }{2\pi ^2 }(2m^2-p^2)\int {\frac{{d\Omega }}{{4\pi }}}\int\limits_0^\Lambda  {d\kappa\,\kappa}\left[ {\frac{1}{{2\kappa a - i\varepsilon }}\frac{1}{{2\kappa b + i\varepsilon }} + \frac{2n(\kappa)}{{2\kappa a - i\varepsilon }}\frac{1}{{2\kappa b + i\varepsilon }}} \right], \label{alpha12}
\end{eqnarray}\noindent where $b = \hat{k}(q+p).$ It is seen from this expression that in the formal limit $T \to 0,$ the second term in the integrand vanishes (because $n(\kappa) \to 0$), leaving us with the vacuum contribution given by the first term. However, this reasoning is not actually legitimate since it is drawn from consideration of the integrals which diverge for $\varepsilon \to 0.$ Taking the limit $T \to 0$ keeping $\varepsilon$ fixed is indeed only formal as it violates condition (\ref{deltacond}). At the same time, it was shown in the preceding section that the finite-temperature effects introduce natural scale for $\varepsilon,$ namely, $ E\sim \alpha mT,$ and this value automatically satisfies (\ref{deltacond}), because $\alpha \ll 1.$ Under this condition, the $\kappa$-integral can be written, using the auxiliary intermediate scale $\lambda,$
\begin{eqnarray}
\int\limits_0^\lambda  {d\kappa\,\kappa}\left[ {\frac{1}{{2\kappa a  - i\varepsilon }}\frac{1}{{2\kappa b  + i\varepsilon }} + \left( {\frac{2}{{\beta \kappa}} - 1} \right)\frac{1}{{2\kappa a  - i\varepsilon }}\frac{1}{{2\kappa b  + i\varepsilon }}} \right] + \frac{1}{4ab}\left[ \int\limits_\lambda ^\Lambda{\frac{d\kappa}{\kappa}  + \int\limits_\lambda ^\infty  \frac{2d\kappa}{\kappa(e^{\beta \kappa}  - 1)}} \right].
\nonumber
\end{eqnarray}
We see that the logarithmic singularity of the vacuum contribution is exactly canceled by the temperature-independent term coming from the expansion $n(\kappa)=1/(\beta \kappa) - 1/2 + O(\beta\kappa).$ The other integrals are evaluated with the help of the formulas
\begin{eqnarray}
&&\int\limits_0^\lambda d\kappa \frac{1}{2\kappa a  - i\varepsilon }\frac{1}{2\kappa b + i\varepsilon} = \frac{\pi}{2\varepsilon}\frac{1}{a  + b} - \frac{1}{4\lambda ab}, \nonumber\\
&&\int\limits_\lambda^\infty\frac{2d\kappa}{\kappa(e^{\beta \kappa}  - 1)}  = \frac{1}{2}\left[\gamma  - \ln (2\pi )\right]  + \frac{2}{{\beta \lambda }} +
\ln(\beta\lambda), \quad \lambda\rightarrow 0 \label{intL}
\end{eqnarray}\noindent ($\gamma$ is the Euler constant).
It remains to set $\varepsilon = -  E,$ and use the formula (\ref{ZnachDelta}).
To this end, it is to be noted that although no restriction on the momentum transfer, $p,$ is imposed {\it a priori}, the fact that Eq.~(\ref{ZnachDelta}) was derived under the condition $|\bm{q}|\ll m$ implies that $p$ also must satisfy $|\bm{p}|\ll m.$ Then $b=a=m,$ and the leading terms in the vertex contribution to $J(p,q)$ take the form
\begin{eqnarray}\label{j2}
J_{(2)}(p,q) = &&\hspace{-0,3cm}
\frac{e^3\gamma^0}{8\pi^2}(2m^2 - p^2)\ln(\beta m) \langle (ab)^{-1}\rangle \nonumber\\&& -
e\gamma^0m\left(2 - \frac{p^2}{m^2}\right)\langle (a+b)^{-1}\rangle \left(1+\frac{e^2}{4\pi^2}\ln(\beta m)\right) = -e\gamma^0 \,.
\end{eqnarray}\noindent

Using the obtained results, it is easy to demonstrate nullification of the Coulomb field of a localized charge. The condition $|\bm{p}| \ll m$ means that the electric field is measured far enough from the charge. More precisely, if the electron is localized near $\bm{x}_0,$ then the distance between this point and the point of observation, $r = |\bm{x} - \bm{x}_0| \gg 1/m.$ Adding the vertex and self-energy contributions gives $J_{(2)} + J_{(3)} = e\gamma^0.$ It follows that the one-loop contribution to the function $J(p,q)$ exactly cancels the tree value (\ref{tree}), so that
\begin{eqnarray}\label{coulomb}
A^{\rm eff}_{0}(x)=0 \quad \text{for} \quad r\gg 1/m.
\end{eqnarray}
\noindent Thus, independently of the particular choice of the electron density matrix, the leading long-range term of its electric potential -- the Coulomb field -- vanishes upon account of the one-loop radiative contributions.

This analysis can be extended to higher orders. For instance, it is not difficult to see that two-loop diagrams contain terms proportional to $(T/\varepsilon)^2,$ and therefore, contribute $O(e)$-terms to the effective potential. In particular, they modify the quantity $E,$ which in turn changes the structure of the $O(e^3)$-terms found at the one-loop level, {\it etc.} Unfortunately, $\varepsilon$-regularization does not allow extension of the above result to all orders, because the mixing of orders in the coupling constant obscures the structure of the perturbation series.

\section{Higher-loop contributions to the effective field}\label{highloop}

\subsection{$\lambda$-regularization}\label{lambdareg}

In addition to mixing the perturbation series, the $\varepsilon$-regularization has the drawback that it breaks the usual factorization of many-loop infrared contributions (see Sec.~\ref{factorization} below). To investigate their structure, we shall now introduce a regularization that does not lead to mixing of various orders of the perturbative expansion, and allows direct summation of the divergent contributions.

First of all, we introduce the usual infrared cutoff $\lambda_0 \ll T,$ so that all loop momenta are now restricted to the interval $\lambda_0 < k < \Lambda.$
Still, this cutoff is insufficient for the purpose of regularizing diagrams with the self-energy insertions. In order to make these and the corresponding counterterm diagrams meaningful, it is necessary to regularize the mass-shell propagators (see Figs.~\ref{Self}, \ref{Kontr}). Leaving the mass shell does not help in this respect, as it precludes factorization of the infrared contributions. By this reason, we introduce the following smearing of the Dirac delta-functions expressing conservation of the 4-momentum in the interaction vertices
\begin{eqnarray}
\delta^4(q_1+k-q_2) \rightarrow \Delta_\lambda(q_1+k-q_2),
\end{eqnarray}\noindent where $\Delta_\lambda(w)$ satisfies
\begin{eqnarray}
&&\int d^4 w \Delta_\lambda(w)=1, \quad \Delta_\lambda(w)=\Delta_\lambda(-w), \\
&&\Delta_\lambda(w\ne 0)  \rightarrow 0 \quad \text{as} \quad \lambda \rightarrow 0.
\end{eqnarray}\noindent A convenient choice is
\begin{eqnarray}
\Delta_\lambda(w) = \frac{1}{\pi^2 \lambda^4}\exp\left({-\frac{w_0^2+\bm{w}^2}{\lambda^2}}\right)\,.
\end{eqnarray}\noindent It will be assumed in what follows that the parameter $\lambda$ characterizing the width of the smeared delta-function satisfies $\lambda \ll \lambda_0.$ We shall also use the following notation
\begin{eqnarray}
\Delta^*_\lambda(w)=\frac{1}{16}\int d^4 \xi \Delta_\lambda\left(\frac{w-\xi}{2}\right)\Delta_\lambda\left(\frac{w+\xi}{2}\right).
\end{eqnarray}

\begin{figure}[h]
\begin{center}
     \includegraphics{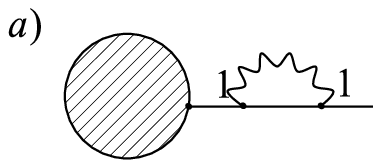}
     \includegraphics{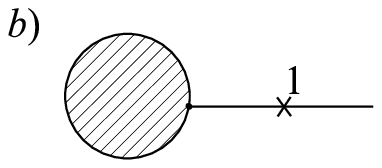}
\caption{Self-energy insertion into external line of a general diagram (a), and the corresponding counterterm diagram (b).}
\label{CKont}
\end{center}
\end{figure}

Next, to ensure convergence of the effective field in the limit $\lambda\to 0,$ we have to introduce counterterms into Lagrangian. To be specific, let us consider diagrams of the type shown in Fig. \ref{CKont}(a). The external line with the self-energy insertion contributes a factor
\begin{eqnarray}&&\label{factorex}
-\int d^4 w_1 \int d^4 w_2 \Delta_\lambda(w_1)\Delta_\lambda(w_2) \frac{1}{2q(w_1+w_2)+i0}(\slashed{q}+\slashed{w_1}+\slashed{w_2}+m)\Sigma^{(11)}(q + w_1) \nonumber\\&& = - \int d^4 w \Delta^*_\lambda(w) \frac{1}{2qw+i0}(\slashed{q}+\slashed{w}+m)\left[\Sigma^{(11)}(q) + O(\lambda)\right].
\end{eqnarray}
Here the quantity $\Sigma^{(11)}(q)$ is taken on the mass shell. The renormalization prescription that the propagator poles be at the physical mass requires vanishing of this quantity, which can be met by introducing a counterterm. The above expression shows that the delta-functions appearing in the two-point counterterm vertices are to be regularized as
\begin{eqnarray}
\delta^4(q_1-q_2) \rightarrow \Delta^*_\lambda(q_1-q_2).
\end{eqnarray}\noindent Then the counterterm diagram shown in Fig. \ref{CKont}(b) reads
\begin{eqnarray}\label{contrDelta}
\int d^4 w \Delta^*_\lambda(w) \frac{1}{2qw+i0}(\slashed{q}+\slashed{w}+m)\Sigma^{(11)}(q).
\end{eqnarray}
The remaining contribution coming from integration of the $O(\lambda)$-term in Eq.~(\ref{factorex}) will be taken care of below when summing all infrared many-loop contributions. It is worth noting that in the limit $p \rightarrow 0,$ the counterterm contributions to the effective field cancel each other, just as in the case of $\varepsilon$-regularization. This can be verified directly repeating the calculation of Sec.~\ref{oneloop} [Cf. Eq.~(\ref{ctcancel})]. However, this property is violated at finite $p$ by the Lorentz-noninvariant integration in Eq.~(\ref{contrDelta}).

\subsection{Factorization of infrared contributions}\label{factorization}

\begin{figure}[h]
\begin{center}
     \includegraphics{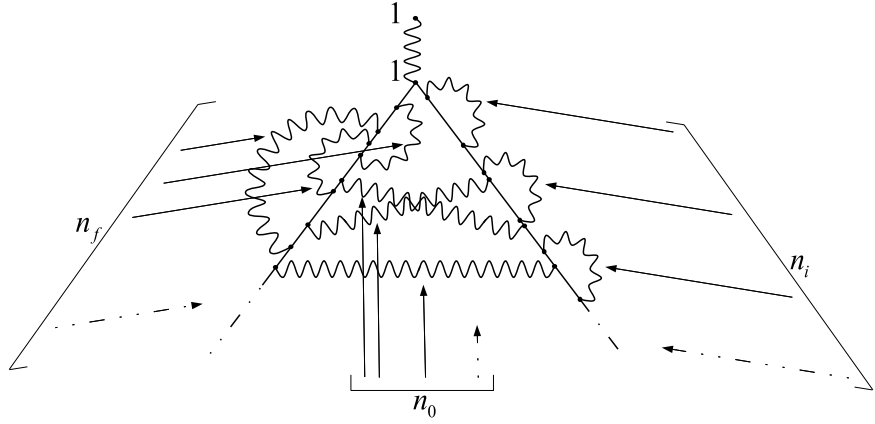}
\caption{General infrared-divergent contribution to the effective field.}
\label{SumA}
\end{center}
\end{figure}
\begin{figure}[h]
\begin{center}
     \includegraphics{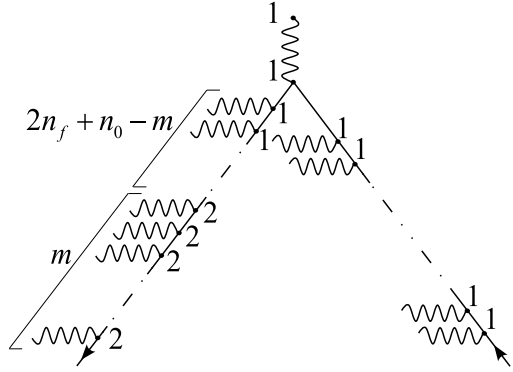}
\caption{Typical diagram involving 2-vertices. It is understood that the horizontal photon lines are arbitrarily paired.}
\label{SumB}
\end{center}
\end{figure}

Now that the finite-momenta contributions to the electron self-energy have been canceled by counterterms as discussed in the preceding section, it remains to take into account  contributions of small virtual photon momenta. The general structure of diagrams to be considered is shown on Fig. \ref{SumA}. In terms of the function $J(p,q),$ the sum of all such diagrams can be written as a series
\begin{eqnarray}\label{series}
J(p,q) = -e\gamma^0 I_{\Lambda}(p,q)\sum_{N=0}^{\infty}(e^2)^N I_N (p,q,\Lambda),
\end{eqnarray}\noindent
where the factor $I_{\Lambda}(p,q)$ corresponds to photons with momenta $k > \Lambda.$ Let us first show that the $\varepsilon$-regularization does not admit the usual factorization of infrared contributions. $I_N$ has the following general form
\begin{eqnarray}\label{I_N}
I_N (p,q,\Lambda)&&=\frac{1}{i^N}\sum_{n_i=0}^N \sum_{n_f=0}^{N-n_i}\int \frac{d^4 k_1}{(2\pi)^4}\cdots\frac{d^4 k_{n_0}}{(2\pi)^4}
m^{2(n_i+n_f)}(m^2-p^2/2)^{n_0} D^{(11)}(k_1)\cdots D^{(11)}(k_{n_0}) \nonumber \\
&&\times \frac{F_{n_i}^{n_0}(q,k_1,\dots,k_{n_0})F_{n_f}^{n_0}(q+p,k_1,\dots,k_{n_0})}{n_i!n_f!n_0!},
\end{eqnarray}\noindent
where $N$ is the number of virtual photon lines, of which $n_i$ ($n_f$) reside on the ingoing (outgoing) electron line, while the remaining $n_0 \equiv N-n_i-n_f$ connect the two electron lines.  In the case of $\varepsilon$-regularization, the function $F_n^{n_0}$ can be conveniently written as
\begin{eqnarray}\label{F_n}
F_{n}^{n_0}(q,k_1,\dots,k_{n_0}) =&&\hspace{-0,5cm}\int d^4 w_1\cdots d^4 w_{2n+n_0}
D^{(11)}(w_2)\cdots D^{(11)}(w_{2n})\nonumber \\
&& \times\delta(k_1-w_{2n+1})\cdots \delta(k_{n_0}-w_{2n+n_0})
\delta(w_2 + w_1)\cdots\delta(w_{2n}+w_{2n-1})\nonumber \\
&& \times
\sum\limits_{\rm perm}\left[ \frac{1}{w_1 q+i\varepsilon} \frac{1}{(w_1+w_2)q+i\varepsilon}\cdots \frac{1}{(w_1+ \cdots +w_{2n+n_0})q+i\varepsilon} \right] \nonumber \\
&& +\text{2-terms},
\end{eqnarray}\noindent
where the sum is over all permutations of indices $1,2,\dots,2n+n_0,$ and ``2-terms'' denotes the contribution of diagrams involving 2-vertices. A partial permutation with respect to the indices 1,2 of the expression in the square brackets yields
\begin{eqnarray}\label{excess}
\left[ \frac{1}{w_1 q+i\varepsilon} \frac{1}{w_2 q+i\varepsilon} \left(1+\frac{i\varepsilon}{(w_1+w_2)q+i\varepsilon}\right)\cdots \frac{1}{(w_1+\cdots +w_{2n+n_0})q+i\varepsilon} \right].
\end{eqnarray}\noindent
It is seen that the usual factorization is hindered by the term proportional to $\varepsilon,$ which is not negligible because $(w_1+w_2)q$ may take on values as small as $\varepsilon.$ Things are different in the $\lambda$-regularization which  replaces Eq.~(\ref{F_n}) by the following
\begin{eqnarray}\label{F_n_Cut}
F_{n}^{n_0}(q,k_1, &&\hspace{-0,3cm} \dots,k_{n_0}) =\int d^4 k_{n_0+1}\cdots d^4 k_{n_0+n} \int d^4 w_1\cdots d^4 w_{2n+n_0}
D^{(11)}(k_{n_0+1})\cdots D^{(11)}(k_{n_0+n}) \nonumber \\
&&\times \Delta(w_{2n+1}-k_1)\cdots \Delta(w_{2n+n_0}-k_{n_0})
\nonumber \\
&& \times \Delta(w_1-k_{n_0+1})\cdots\Delta(w_{2n-1}-k_{n_0+n}) \Delta(w_2+k_{n_0+1})\cdots\Delta(w_{2n}+k_{n_0+n})
\nonumber \\
&& \times \sum\limits_{\rm perm}\left[ \frac{1}{w_1 q+i\varepsilon} \frac{1}{(w_1+w_2)q+i\varepsilon}\cdots\frac{1}{(w_1+ \cdots +w_{2n+n_0})q+i\varepsilon} \right] + \text{2-terms}.
\end{eqnarray}\noindent
Now the second term in the large parentheses in Eq.~(\ref{excess}) is $O(\varepsilon/\lambda_0),$ and it is not difficult to check that so are all other terms proportional to $\varepsilon,$ appearing in the course of the permutation. Therefore, under the assumption $m\lambda_0 \gg \varepsilon,$ they can be omitted, and we may consistently set $\varepsilon=0.$

Next, let us show that the 2-terms do not contribute to the effective field in the $\lambda$-regularization. Figure \ref{SumB} depicts general structure of the corresponding diagrams. By the reason explained at the beginning of Sec.~\ref{oneloop}, diagrams with a 1-vertex appearing to the left of a 2-vertex vanish. If a diagram has $m > 1$ vertices of the 2-type on the outgoing electron line, then the corresponding sum over permutations reads
\begin{eqnarray}&&
\hspace{-0,3cm}\sum\limits_{\rm perm}\left[ \frac{1}{w_1 q-i0}\cdots \frac{1}{(w_1 +\cdots +w_{m-1}) q-i0}
\left(\frac{1}{(w_1+\cdots +w_m) q+i0}\right. \right. \nonumber \\
&&\left. \left. \hspace{-0,3cm}-\frac{1}{(w_1 +\cdots +w_m) q-i0}\right) \frac{1}{(w_1 +\cdots +w_{m+1}) q+i0}
\dots \frac{1}{(w_1+\cdots +w_{2n_f+n_0})q+i0}  \right]=\nonumber \\&&
\hspace{-0,3cm}=\sum\limits_{\rm perm}\frac{1}{m!}\left[ \left(\frac{1}{w_1 q-i0}\cdots\frac{1}{w_m q-i0} - \frac{1}{w_1 q-i0}\cdots\frac{1}{w_m q-i0}\right)\frac{1}{(w_1 +\cdots +w_{m+1}) q+i0}\cdots\right]\nonumber\\&& \hspace{-0,3cm} =0,\nonumber
\end{eqnarray}\noindent
where we explicitly performed the permutation over indices $1,...,m,$ and used the factorization rule proved above. In the case $m=1,$ expression in the square brackets takes the form
\begin{eqnarray}
\left(\frac{1}{w_1 q+i0}-\frac{1}{w_1 q-i0}\right)\frac{1}{(w_1+w_2) q+i0}\cdots\frac{1}{(w_1 +\cdots +w_{2n_f+n_0}) q+i0}. \nonumber
\end{eqnarray}\noindent
This vanishes, too, because $w_i \neq 0$ in the $\lambda$-regularization. Thus, the contribution of the $\text{2-terms}$ is zero. This result is discussed from a more general point of view in the Appendix.

We are ready to perform summation of the infrared contributions in the series (\ref{series}). This is very similar to the standard treatment of the loop infrared divergences (see, {\it e.g.}, \cite{Weinberg}). The sum over permutations in Eq.~(\ref{F_n_Cut}) becomes
\begin{eqnarray}\label{FnFact}
\frac{1}{w_1 q+i0} \frac{1}{w_2 q+i0}\cdots \frac{1}{w_{2n+n_0}q+i0}\,.
\end{eqnarray}\noindent
Substituting Eq.~(\ref{F_n_Cut}) into Eq.~(\ref{I_N}), and designating
\begin{eqnarray}
d_{st}=-i q_s q_t \int \frac{d^4 k}{(2\pi)^4} \int d^4 w_1 \int d^4 w_2
\frac{\Delta(w_1-\eta_s k)\Delta(w_2+\eta_t k) D^{(11)}(k)}{(w_1 q_s+i0)(w_2 q_t+i0)}, \nonumber
\end{eqnarray}\noindent
where $s,t=1,2$, $q_1=q$, $q_2=q+p$, $\eta_1=1$, $\eta_2=-1$, we obtain
\begin{eqnarray}\label{inpq}
I_N (p,q,\Lambda)=\sum_{n_i=0}^N \sum_{n_f=0}^{N-n_i}
\frac{d_{11}^{n_i}}{n_i!\,2^{n_i}} \frac{d_{22}^{n_f}}{n_f!\,2^{n_f}} \frac{d_{12}^{n_0}}{n_0!}=
\frac{(d_{11}+d_{22}+2d_{12})^N}{N!\,2^N}
\end{eqnarray}\noindent
Under the assumption $\lambda\ll\lambda_0,$ $d_{st}$ can be evaluated as
\begin{eqnarray}\label{dij}
d_{st}&=&i\eta_s \eta_t q_s q_t \int \frac{d^4 k}{(2\pi)^4} \int d^4 w_1 \int d^4 w_2 \frac{\Delta(w_1)\Delta(w_2)D^{(11)}(k)}{(k q_s)(k q_t)} + O\left(\frac{\lambda}{\lambda_0}\right).
\end{eqnarray}\noindent Therefore, in the limit of removed $\lambda$-regularization,  \begin{eqnarray}
d_{st} = i\eta_s \eta_tq_s q_t
\int \frac{d^4 k}{(2\pi)^4} \frac{D^{(11)}(k)}{(k q_s)(k q_t)}\nonumber
\end{eqnarray}\noindent
Substitution of Eq.~(\ref{inpq}) into Eq.~(\ref{series}) gives
\begin{eqnarray}\label{Jpq}
J(p,q) = -e\gamma^0\exp\left\{\frac{e^2}{2}(d_{11}+d_{22}+2d_{12})\right\} I_{\Lambda}(p,q)
\end{eqnarray}\noindent
The integrals $d_{st}$ can be computed in the same way as the $J_{(2)}$-integral in Sec.~\ref{vertex}. One has
\begin{eqnarray}
d_{st}=\frac{\eta_s \eta_t}{4\pi^2} q_s q_t
\int \frac{d\Omega}{4\pi}\frac{1}{(\hat{k}q_s)(\hat{k}q_t)}
\int\limits_{\lambda_0}^\Lambda \frac{d\kappa}{\kappa}
\left[1+n(\bm{k}) \right],\nonumber
\end{eqnarray}\noindent
and the use of Eq.~(\ref{intL}) brings this function to the form
\begin{eqnarray}
d_{st}=\frac{\eta_s \eta_t}{4\pi^2} q_s q_t
\int \frac{d\Omega}{4\pi}\frac{1}{(\hat{k}q_s)(\hat{k}q_t)}
\left[\frac{2}{\beta \lambda_0}+\ln(\beta \Lambda)  + \left(\gamma  - \ln (2\pi )\right)\right].\nonumber
\end{eqnarray}\noindent
Thus, retaining the singular terms in the exponent gives finally
\begin{eqnarray}\label{jfinal}
J(p,q) = -e\gamma^0\exp\left\{\frac{\alpha}{\pi}\left[1-\left(m^2-p^2/2\right)
\langle (a b)^{-1}\rangle\right]\left[\frac{2}{\beta \lambda_0}+\ln(\beta \Lambda) \right]\right\}I_{\Lambda}(p,q)\,.
\end{eqnarray}\noindent
The expression in the exponent is negative for all $p.$ For small momentum transfer, for instance, one has
\begin{eqnarray}
J(p,q) = -e\gamma^0\exp\left\{- \frac{\alpha \bm{p}^2}{3\pi m^2}\left[\frac{2}{\beta \lambda_0}+\ln(\beta \Lambda) \right]\right\}I_{\Lambda}(p,q)\,, \quad |\bm{p}|\ll m.\nonumber
\end{eqnarray}\noindent
As was already mentioned in the Introduction, the question as to whether the infrared divergence of the effective field matters at large distances can be resolved only after summing up the infinite series of local terms proportional to the $\delta$-function derivatives. We now see that the soft-photon contribution leads to vanishing of the effective electric field at all distances from the charge: substituting the above expression for $J(p,q)$ into Eq.~(\ref{Fur_e}), and taking the limit $\lambda_0\rightarrow 0$ yields
$$ A^{\rm eff}_{0}(x)=0 \quad \text{for all} \quad \bm{x}.$$

\section{Discussion and conclusions}\label{conclude}

We have investigated the effective electric potential of a localized electron using two different regularizations of the infrared divergences. In both cases, our results suggest that the effective field calculated using the perturbation theory vanishes. Though leading to the same conclusion, the two regularization schemes are quite different. While vanishing of the field in the $\lambda$-regularization resembles that of the $S$-matrix amplitudes with a fixed number of real photons, and occurs as the result of summation of all divergent diagrams in the infinite perturbation series, the field nullification in the $\varepsilon$-regularization takes place already at the one-loop level. To the best of our knowledge, nothing similar has been encountered in calculations of other field-theoretic expectation values. In addition to that, this scheme reveals a number of important cancelations between various diagrams, described in Sec.~\ref{oneloop}.

It should be stressed, however, that our consideration is inherently perturbative, and hence is not complete, because the power-expansion with respect to the coupling constant breaks down, explicitly or implicitly, in both schemes. As to the $\varepsilon$-regularization, we found that this scheme is quite natural in view of the fact that the finite-temperature effects provide an effective infrared regulator proportional to the photon bath temperature, which renders any diagram with a given number of loops infrared-finite. But the resulting mixing of orders in $\alpha$ makes it necessary to take into account diagrams with arbitrary number of loops. This complication bears some resemblance to that caused by the gluon mass generation in high-temperature QCD \cite{linde}, the difference being that in our case, it is imaginary contribution to the mass that is responsible for the breakdown of the perturbative expansion.\footnote{That the thermal field theory in its present formulation is inherently non-perturbative was discussed from a more general standpoint in Ref.~\cite{landsman1991}.}

On the other hand, as we showed in Sec.~\ref{highloop}, the $\lambda$-regularization is free of this drawback. This scheme allowed us to consistently take the limit $\varepsilon\to 0$ in the cut integrals, and rendered diagrams with 2-vertices vanishing. Specifically, the assumption $\lambda_0 \gg |\varepsilon|/m$ permitted us to overcome the question about generation of an effective $\varepsilon,$ which is formally zero in this regularization. In itself, this does not contradict the fact that it is nonzero in the other scheme, because, as was mentioned above, the perturbative calculation using the $\varepsilon$-regularization is not complete. However, appearance of an imaginary part in the particle mass seems to be a general property of thermal field theory, signifying the presence of a relaxation mechanism driving the quantum field to the equilibrium with the heat bath \cite{weldon1983,semenoff1985,keil1989}. Therefore, if the result $E\ne 0$ is scheme-independent, the limit $\lambda_0 \to 0$ cannot be taken in Eq.~(\ref{jfinal}), because in effect, it would violate the assumption $\lambda_0 \gg |\varepsilon|/m.$ Rather, Eq.~(\ref{jfinal}) can be used in this case to estimate the function $J(p,q).$ Indeed, the only formal condition that the parameter $\lambda_0$ must  satisfy is $\lambda_0 \ll T$ (see Sec.~\ref{lambdareg}). At the same time, $E/m \sim \alpha T \ll T,$ so that a rough estimate for $J(p,q)$ can be obtained by replacing $\lambda_0 \to E/m.$ Equation (\ref{jfinal}) then would give
$J \sim - e\gamma^0\exp\left\{O(\alpha^0)\right\}I_{\Lambda},$ showing that if $E\ne 0,$ the consideration based on the $\lambda$-regularization goes beyond the perturbation theory, too.

We would like to emphasize that this conclusion about essentially non-perturbative nature of the infrared problem is directly related to the question of whether the infrared divergences affect the long-range behavior of the effective field. We have seen that despite the infrared-divergent terms in the effective field are local at every loop order, the sum of the infinite number of such terms is not (see Sec.~\ref{factorization}). Specifically, if the limit $\lambda_0\to 0$ can actually be taken in Eq.~(\ref{jfinal}), the effective field vanishes at all distances from the charge. On the other hand, if the temperature effects set a lower $O(\alpha)$-bound on $\lambda_0,$ then the infrared factor in $J(p,q)$ corrects the effective field at large distances; though of the relative order $1/r^2,$ this correction cannot be determined within the perturbation theory.

\acknowledgements{We thank Drs. P.~Pronin, G.~Sardanashvili, and especially A.~Baurov (Moscow State University) for their continuing interest to our work and useful discussions.}

\begin{appendix}

\section*{Appendix: Ward identities and a relation between 2-terms and $E.$}
\begin{figure}[h]
\begin{center}
    \includegraphics{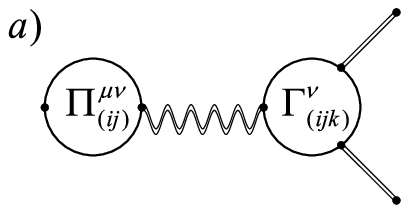}
    \includegraphics{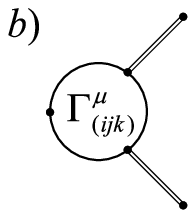}
\caption{General structure of diagrams representing the function $G^{\mu}_{(ijk)}.$ The blobs denote irreducible parts -- the photon polarization operator, $\Pi,$ and the proper electron-photon vertex, $\Gamma.$ Double lines are the exact propagators.}
\label{WardFig}
\end{center}
\end{figure}

As we have seen, the perturbation expansion in powers of the coupling constant breaks down in the case $E\ne 0.$ In view of this breakdown, it is natural to look for non-perturbative means for investigating the infrared problem. One of such means is the use of the Ward identities. Unfortunately, the introduction of $\varepsilon$-regulator, necessary in the real-time techniques to deal with the products of singular functions, violates the usual Ward identities. Still, consideration of the unregularized identities makes sense by the following reason. As we have seen, the finite $E$ plays the role of an effective $\varepsilon$-regulator, though it necessitates analytical continuation to the negative values of $\varepsilon.$ Therefore, under the assumption that the exact electron self-energy admits this continuation, the existence of $E\ne 0$ suggests that $\varepsilon$ can be consistently set equal to zero.

To derive the Ward identities for our model, let us consider the following exact Green function
\begin{eqnarray}
G^{\mu}_{(ijk)}(x,y,z) = {\rm Tr} \left( T_c\left[ \psi_i(z) J_j^\mu(x) \bar{\psi}_k(y)\right]e^{- \beta H_\phi } \varrho\right)\,,
\end{eqnarray}\noindent
where $J^\mu(x)={\bar \psi}(x)\gamma^{\mu}\psi(x)$ is the electromagnetic current, and the subscript indices indicate position of the corresponding time argument on the contour $C$ (see Sec.~\ref{prelim}). As before, the components relevant to the infrared properties of the effective field have $j=k=1$ (other components can be dealt with using complex conjugation). Using the current conservation $\partial_\mu J^\mu(x)=0,$ we find for their 4-divergences
\begin{eqnarray}\label{conserv1}
\frac{\partial}{\partial x^\mu}{\rm Tr} \left( T_c\left[ \psi_1(z) J_1^\mu(x) \bar{\psi}_1 (y)\right]e^{- \beta H_\phi } \varrho\right) =
\delta^4(x-y) {\rm Tr} \left( T_c\left[ \psi_1(z) \bar{\psi}_1(x)\right]e^{- \beta  H_\phi} \varrho\right) \nonumber \\
-\delta^4(x-z) {\rm Tr} \left( T_c\left[ \psi_1(x) \bar{\psi}_1 (y)\right]e^{- \beta H_\phi} \varrho\right)\,,
\end{eqnarray}\noindent
\begin{eqnarray}\label{conserv2}
\frac{\partial}{\partial x^\mu}{\rm Tr} \left( T_c\left[ \psi_2(z) J_1^\mu(x) \bar{\psi}_1 (y)\right]e^{- \beta H_\phi} \varrho\right)=
\delta^4(x-y) {\rm Tr} \left( T_c\left[ \psi_2(z) \bar{\psi}_1 (x)\right]e^{- \beta  H_\phi}\varrho\right)\,.
\end{eqnarray}\noindent
Diagrams representing $G^{\mu}_{(ijk)}$ are shown schematically in Fig.~\ref{WardFig}. To compute the left hand sides of Eqs.~(\ref{conserv1}), (\ref{conserv2}), we shall use the Ward identity for the photon polarization operator $\Pi^{\mu\nu}(x-y),$ \cite{landsman}
$$\partial_\mu \Pi^{\mu\nu}_{ij}(x-y)=0.$$ It follows from this identity that diagrams in Fig.~\ref{WardFig}(a) do not contribute to the left hand sides of Eqs.~(\ref{conserv1}), (\ref{conserv2}). Going over to the momentum space, these equations thus yield
\begin{eqnarray}\label{ConsImpuls1}
\EuScript{D}_e^{(11)}(q+p)p_\mu \Gamma^\mu_{(111)}(q+p,q)\EuScript{D}_e^{(11)}(q) = \EuScript{D}_e^{(11)}(q+p)-\EuScript{D}_e^{(11)}(q),
\end{eqnarray}\noindent
\begin{eqnarray}\label{ConsImpuls2}
p_\mu\left[\EuScript{D}_e^{(21)}(q+p) \Gamma^\mu_{(111)}(q+p,q)+\EuScript{D}_e^{(22)}(q+p)\Gamma^\mu_{(211)}(q+p,q) \right]\EuScript{D}_e^{(11)}(q) = \EuScript{D}_e^{(21)}(q+p),
\end{eqnarray}\noindent
where $\EuScript{D}_e^{(ij)}$ and $\Gamma^\mu_{(i11)}$ are the exact electron propagators and proper vertices. Notice that all other vertex components can be derived from these by complex conjugation. The first of these identities can be rewritten as
\begin{eqnarray}
p_\mu \Gamma^\mu_{(111)}(q+p,q) = (\EuScript{D}_e^{(11)}(q))^{-1}-(\EuScript{D}_e^{(11)}(q+p))^{-1}.\nonumber
\end{eqnarray}\noindent
Using this in Eq.~(\ref{ConsImpuls2}), and taking into account that  $\EuScript{D}_e^{(21)}(q) = \EuScript{D}_e^{(11)}(q) + \EuScript{D}_e^{(22)}(q)$ for $q_0>0,$ we bring the other Ward identity to the form
\begin{eqnarray}
p_\mu \Gamma^\mu_{(211)}(q+p,q) = (\EuScript{D}_e^{(22)}(q+p))^{-1} + (\EuScript{D}_e^{(11)}(q+p))^{-1}.\nonumber
\end{eqnarray}\noindent This identity is of particular importance as it gives insight into the connection between the appearance of $E\ne 0$ and the vanishing of 2-terms, underlying factorization of the infrared contributions to the effective field. Namely, rewriting it as
\begin{eqnarray}
p_\mu \Gamma^\mu_{(211)}(q+p,q) = 2i{\rm Im}\Sigma_F(q+p),\nonumber
\end{eqnarray}\noindent we see that $\Gamma^\mu_{(211)}(q+p)$ cannot vanish, if $E\ne 0$ (in fact, the exact vertex is singular for $p\to 0$ in this case). Accordingly, this function was found non-vanishing in the one-loop calculation of Sec.~\ref{vertex}. On the other hand, it was demonstrated in Sec.~\ref{factorization} that using the $\lambda$-regularization, one can consistently set $\varepsilon = 0$ at any given loop order, and prove that the sum of diagrams involving 2-vertices vanishes. Accordingly, $E$ is formally zero in the $\lambda$-regularized perturbation theory. As discussed in Sec.~\ref{conclude}, if $E$ does not vanish in the complete theory, this result is valid only for $\lambda_0\gg E/m.$

\end{appendix}


\begin{thebibliography}{}

\bibitem{equiv}
E.~S.~Fradkin, Trudy FIAN {\bf 29}, 7 (1965);
R.~Fukuda and T.~Kugo, Phys.~Rev. D {\bf 13}, 3469 (1976);
R.~Kallosh and I.~V.~Tyutin, Yad. Fiz. {\bf 17}, 190 (1973)
[Engl. Trans. Sov.~J.~Nucl.~Phys. {\bf 17}, 98 (1973)]; P.~M.~Lavrov, I.~V.~Tyutin, and B.~L.~Voronov, Yad.~Fiz. {\bf 36}, 498 (1982) [Engl. Transl. Sov.~J.~Nucl.~Phys. {\bf 36}, 292 (1982)].

\bibitem{bohr1}
N.~Bohr and L.~Rosenfeld, Kgl.~Danske~Vidensk.~Selskab.,
Math.-Fys.~Medd. {\bf 12}, 3 (1933).

\bibitem{bohr2}
N.~Bohr and L.~Rosenfeld,  Phys.~Rev. {\bf 78}, 794 (1950).

\bibitem{dewitt}
B.~S.~DeWitt, The quantization of geometry, in: {\it Gravitation:
an introduction to current research}, ed.: L. Witten (Wiley, New-York, 1962) p. 266.

\bibitem{bloch}
F.~Bloch and A.~Nordsieck, Phys.~Rev. {\bf 37}, 54 (1937); D.R. Yennie, S.C. Frautschi, and H. Suura, Ann. Phys. (NY), {\bf 13}, 379 (1961); S.~Weinberg, Phys.~Rev. {\bf 140}, B515 (1965); G.~Grammer and D.~R.~Yennie, Phys.~Rev. {\bf D8}, 4332 (1973).

\bibitem{leenaum}
T.~Kinoshita, J.~Math.~Phys. {\bf 3}, 650 (1962); T.~D.~Lee and M.~Nauenberg, Phys.~Rev. {\bf 133}, B1549 (1964); G.~Sterman and S.~Weinberg, Phys.~Rev.~Lett. {\bf 39}, 1416 (1977).

\bibitem{donoghue}
J.~F.~Donoghue, Phys.~Rev.~Lett. {\bf 72}, 2996 (1994);
Phys.~Rev.~D {\bf 50}, 3874 (1994).

\bibitem{toyoda}
T.~Toyoda, Phys.~Rev. {\bf 77}, 353 (1950).

\bibitem{nambu}
Y.~Nambu, Prog.~Theor.~Phys. {\bf 5}, 614 (1950).

\bibitem{kikugawa}
K.~Hiida, M.~Kikugawa, Prog.~Theor.~Phys. {\bf 46}, 1610 (1971);
K.~Hiida, H.~Okamura, Prog.~Theor.~Phys. {\bf 47}, 1743 (1972);
H.~Okamura {\it et al.}, Prog.~Theor.~Phys. {\bf 50}, 2066 (1973);
K.~Yokoya {\it et al.}, Phys.~Rev.~D {\bf 16}, 2655 (1977).

\bibitem{serber}
R.~Serber, Phys.~Rev.~{\bf 48}, 49 (1935); A.~E.~Uehling, Phys.~Rev.~{\bf 48}, 55 (1935).

\bibitem{bialb}
I.~Bialynicki-Birula and Z.~Bialynicka-Birula, {\it Quantum Electrodynamics} (Oxford, Pergamon, 1975); J.~Z.~Kaminski, J.~Phys.~A: Math.~Gen. {\bf 16}, 2587 (1983).

\bibitem{tryon}
E.~P.~Tryon, Phys.~Rev.~Lett. {\bf 32}, 1139 (1975).

\bibitem{donoghue1985}
J.~F.~Donoghue, B.~R.~Holstein, and R.~W.~Robinett, Ann.~Phys. {\bf 164}, 233 (1985).

\bibitem{altherr}
T.~Altherr, Phys.~Lett. {\bf 262B}, 314 (1991).

\bibitem{manjavidze}
J.~Manjavidze, arXiv:hep-ph/9510251.

\bibitem{weldon}
H.~A.~Weldon, Phys.~Rev. D {\bf 49}, 1579 (1994); Nucl.~Phys. {\bf A566}, 581c (1994).

\bibitem{indumathi}
D.~Indumathi, Ann.~Phys. {\bf 263}, 310 (1998); arXiv:hep-ph/9607206.

\bibitem{muller}
A.~Muller, arXiv:hep-th/9912240.

\bibitem{landsman}
N.~P.~Landsman and Ch.~G.~van Weert, Phys.~Rep. {\bf 145}, 141 (1987).

\bibitem{Keldysh}
L.~V.~Keldysh, Sov.~Phys.~JETP {\bf 20}, 1018 (1964).

\bibitem{Schwinger}
J.~Schwinger, J.~Math.~Phys. {\bf 2}, 407 (1961).

\bibitem{niemi}
A.~J.~Niemi and G.~W.~Semenoff, Ann.~Phys. {\bf 152}, 105 (1984);
Nucl.~Phys. {\bf B230} [FS10], 181 (1984).

\bibitem{landsman1991}
N.~P.~Landsman, Nucl.~Phys. {\bf A525}, 397c (1991).

\bibitem{semenoff1985}
R.~L.~Kobes and G.~W.~Semenoff, Nucl.~Phys. {\bf B260}, 714 (1985); {\bf B272}, 329 (1986).

\bibitem{keil1989}
W.~Keil, Phys.~Rev. D {\bf 40}, 1176 (1989).

\bibitem{Weinberg}
S.~Weinberg, {\it The Quantum Theory of Fields} (Cambridge Univ. Press, 1995) Vol.1, Ch.13.

\bibitem{linde}
A.~Linde, Phys.~Lett. {\bf 93B}, 327 (1980); D.~J.~Gross, R.~D.~Pisarski, and L.~G.~Yaffe, Rev.~Mod.~Phys. {\bf 53}, 43 (1981).

\bibitem{weldon1983}
H.~A.~Weldon, Phys.~Rev. D {\bf 28}, 2007 (1983).


\end{thebibliography}
\end{document}